\documentstyle[twocolumn,prb,aps]{revtex}
\tightenlines
\begin{document}
\draft
\title{Parastatistics of charged bosons partly localised by impurities}
\author{A.S. Alexandrov \and R.T. Giles}
\address{Department of Physics, Loughborough University, Loughborough, LE11
3TU, U.K.}
\date{\today}
\maketitle
\begin{abstract}
Coulomb repulsion  is taken into account
to derive the thermodynamics of charged bosons in a random external
potential.
 A simple
analytical form of the partition function is proposed
 for the case of  non-overlapping localised states (i.e. a small
amount of disorder).
 The density of
 localised bosons and the specific heat show a peculiar non-uniform
 temperature dependence  below the Bose-Einstein condensation temperature.
 The superfluid - Bose-glass phase diagram is discussed. A new phase is
 predicted which is a Bose-glass at $T=0$ but a superfluid at finite $T$.
\end{abstract}

\section{Introduction}

The charged Bose-gas (CBG), being an important reference
system of quantum statistics, has become of particular physical
interest in the field of
high temperature superconductivity \cite{alemot}.
A long time ago Schafroth \cite{sha}  demonstrated that an ideal gas of
charged bosons
exhibits the Meissner-Ochsenfeld effect
 below the ideal Bose-gas condensation temperature. Later, the
one-particle excitation
spectrum at $T=0$ was calculated by Foldy \cite{fol} who worked at zero
temperature using the Bogoliubov \cite{bog}
approach. The Bogoliubov method leads to the result that
 the elementary excitations of the system have, for small
momenta, energies characteristic of plasma oscillations which pass over
smoothly for large momenta to the energies characteristic of single
particle excitations. Further investigations have been carried out at or
near $T_{c}$, the transition temperature for the gas; these have
been concerned with the critical exponents \cite{bis} and
the change in the transition temperature from that of the ideal gas
\cite{fet,bis}.
The RPA dielectric response function  and
  screening in a CBG have been
studied in the high-density limit \cite{hor,fra}.
 The theory of the CBG beyond the lowest order Bogoliubov
 approximation  was discussed by Lee and Feenberg \cite{lee}
  and by Brueckner\cite{bru}, both of whom obtained the next order
correction to the ground state energy.
 Woo and Ma \cite{woo} calculated numerically the correction to the
Bogoliubov
excitation spectrum. Alexandrov \cite{ale2} found the critical
magnetic field $H_{c2}(T)$ at which the CBG is condensed; the predicted
temperature dependence of $H_{c2}$ was observed both in low-$T_{c}$
and high-$T_{c}$ cuprates, in which
the coherence volume
estimated from the heat capacity measurements  is
comparable with or even less than the unit cell volume.

The doped Mott insulators are intrinsically disordered. Hence, the
localisation of
carriers in a random potential plays a crucial role in their
low-temperature thermodynamics and transport properties.
 As an example, if a fraction of  bipolarons is localised by
disorder and the Coulomb repulsion in localised states is sufficiently
strong, then the number
of delocalised bosons is
proportional to $T$ while the boson-boson inelastic scattering rate  is
proportional to $T^{2}$;  this can explain both  the
linear temperature dependence of the in-plane resistivity and the Hall
density observed in the cuprates \cite{alemot}. The  picture of
interacting bosons with a short-range
interaction
filling up all localised single-particle states in a random potential
and Bose-condensing into the first extended state is  known in the
literature \cite{her,ma,fis}.
To calculate the density of localised bosons  one has to
 take into account  the repulsion
between them.  One cannot ignore the fact that the
localisation length $\xi$ generally varies with energy and diverges at
the mobility edge. One would expect that the number of $hard$ $core$
bosons
in a localised state near the mobility edge diverges in a similar way to
the localisation length. Therefore only a repulsive interaction can
stop all particles condensing into the lowest localised state. Thus, as
stressed by Fisher $et$ $al$ \cite{fis}, there is no sensible
non-interacting starting point in the Bose gas about which to perturb (in
contrast to the Fermi gas).

The scaling analysis of neutral
bosons in  random potentials by Fisher $et$ $al$ \cite{fis} 
describes  the Bose-glass
- superfluid
phase transition at zero temperature with the increasing density and (or)
the hopping strength.  However the analysis is limited to the critical
region near the transition and predicts \cite{Fisher90} the universal
resistance $R^{*}$ at the Mott insulator to superconductor transition (in
two dimensions) while data \cite{haviland} on amorphous films suggest a wide
range of $R^*$.  Simple analytically solvable models of interacting bosons
in a random potential might therefore be helpful.  Lee and Gunn \cite{Gunn}
proposed a picture of neutral bosons in which the ``true'' extended Bose
condensate co-exists with bosons in localised states.  In this paper we
develop a similar model for charged bosons interacting via Coulomb forces.

The number of $charged$ bosons in a single potential well is determined by
the competition
between the long-range Coulomb potential energy
$\simeq{4e^{2}/\epsilon_{0}\xi} $ and
the binding
energy $E_{c}-E$, where $\epsilon_{0}$ is the background dielectric
constant and $E_{c}$ is the mobility edge. If the Coulomb interaction is
strong and the
localisation length~$\xi$ diverges as $\xi\propto{(E_{c}-E)^{-\nu}}$ with
$\nu<1$, then each potential well cannot contain more than one
boson\cite{alemot}.
 Within this approximation the localised bosons obey Fermi-Dirac
 statistics because Coulomb repulsion has the same effect as the
 Pauli exclusion principle. In the extreme case of the hydrogen atom the
average electron-nucleus distance is inversely proportional to the binding
energy, i.e. $\nu=1$. The hydrogen negative ion exists but
with rather low electron affinity   so the doubly
charged negative ion
$H^{2-}$ exists only as a resonant state. However, in general, the
exponent  $\nu$ can be larger then unity and the Coulomb repulsion
is
not infinite, so the statistics of localised
charged bosons is neither Bose nor Fermi.

In this paper we study the parastatistics of partly localised charged bosons
in the
 superfluid phase. In this phase the chemical potential
$\mu$ is exactly at the mobility edge, $\mu=E_{c}$. We assume that the
density of bosons $n$ is not very high or the dimensionality is reduced so
the  localised
states are not completely screened out (as in the case of the Mott
transition in semiconductors).
For a $3D$ system that means that  the  dimensionless strength of the
Coulomb repulsion should be about unity or larger,
$r_{s}=4me^{2}/\epsilon_{0}
(4\pi n/3)^{1/3}\geq 1$ where $m$ is the boson mass and we take $\hbar=1$. At
the same time  the density should  not be extremely low if  delocalised
bosons are in a superfluid phase. A rather large magnitude of $r_{s}>>1$
is
necessary for the Wigner crystallisation of charged carriers. Therefore,
the above condition takes place practically in the whole relevant region
of the carrier density. Moreover in reduced dimensions, relevant for cuprates,
the localised states cannot be screened out even if $r_{s} \leq 1$, and for
$r_{s}>1$ the localisation length $\xi$ is essentially unaffected by
delocalised bosons or the Bogoliubov collective mode.  Hence our picture
suggests that some particles remain in localised states while others are in
the ``true'' extended Bose condensate; for neutral bosons this picture has
been justified by Lee and Gunn \cite{Gunn}.

\section{Partition function of localised charged bosons}

The  Hamiltonian  of charged bosons  on an
oppositely charged background (to ensure charge neutrality) in an external
random field  with the  potential $U({\bf r})$ is given by

\begin{eqnarray}
H&=&\int d{\bf r} \psi^\dagger({\bf r})\left[-{{\bf \nabla}^2
\over{2m}}
-\mu+U({\bf r})\right]\psi({\bf r})\cr
&+&
\frac 1 2 \int d{\bf r} \int d{\bf r'} V({\bf r} -
{\bf r'})\psi^\dagger({\bf r})
\psi({\bf r})\psi^\dagger({\bf r'})\psi({\bf r'}).
\end{eqnarray}
The Fourier component of the Coulomb potential for bosonic charge $2e$
is $ V({\bf k})= 16\pi {e}^2/\epsilon_{0} k^{2}$
  in a $3D$ system, and $V({\bf k})= 8\pi
{e}^2/\epsilon_{0} k$
in a $2D$ system with a three dimensional interaction. For a low amount of
disorder a single particle spectrum consists
 of  localised discrete levels below the bottom of the conduction band
 $E_{c}$.  At some
 finite temperature, $T_{c}$, bosons are  condensed at $E=E_{c}$ so that
 $\mu=E_{c}$. If the Coulomb repulsion is strong one can expect that
each localised state below $E_{c}$ is occupied by one or a few bosons.
 The excitation spectrum of the delocalised charged superfluid has a gap of
the order
of the plasma frequency \cite{fol}. Therefore the low-temperature
thermodynamics is controlled by the excitation of the shallow localised
states while the Bogoliubov collective modes can be ignored. In $3D$
their contribution is exponentially small while in $2D$ their energy
scales as $T^{5}$ and the specific heat as $T^{4}$. Even in the
case of a short range repulsion the sound modes yield an energy
proportional to $T^{d+1}$  and hence a specific heat which behaves
like $C\propto T^{d}$ (where $d$ stands for the dimensionality) \cite{fis}.
Therefore
the contribution to thermodynamics from the delocalised bosons appears to be
negligible at low temperatures compared with that from bosons localised in
shallow potential wells for $d\geq 2$. So in the following we
calculate the partition function and specific heat of  localised
bosons only.

When two or more charged bosons
are in a single localised state of energy $E$ there may be significant
Coulomb energy
and we try to take this into account as follows.  The localisation length
$\xi$ is assumed to depend on $E$ via
\begin{equation}
  \xi\sim \frac{1}{\left(-E\right)^\nu},
  \end{equation}
  where $\nu>0$. The Coulomb energy of $p$ charged bosons confined within a
radius $\xi$ can be expected to be of order
\begin{equation}
   \frac{p(p-1)e^2}{\epsilon_0 \xi}.
\end{equation}
Thus the total energy~$w$ of $p$ bosons in a localised state of energy
$E$ is taken to be
\begin{equation}
  w=pE + p(p-1)\kappa\left(-E\right)^\nu
\end{equation}
where $\kappa>0$. Hence, as mentioned above, we see that the behaviour of
charged bosons in
localised states can be thought of as intermediate between Bose-Einstein
statistics and Fermi-Dirac statistics.  When $\kappa=0$ we have an equally
spaced set of levels, i.e. Bose-Einstein behaviour, whereas when
$\kappa=\infty$ we have Fermi-Dirac behaviour since the only levels with
finite energy are $p=0$ and $p=1$, and thus an exclusion principle is
enforced.  When $0<\kappa<\infty$ we have the intermediate `parastatistics'
that
the level spacing $\delta w\rightarrow\infty$ as $p\rightarrow\infty$.  This
behaviour is closely analagous to that observed in the Coulomb blockade
model which applies to quantum dots \cite{Kastner}.

We are primarily interested in the properties of the superconducting
phase, namely the phase in which a true (extended) Bose condensate is
present. The only
state in which the true  condensate can form is the lowest energy delocalised
state $E_{c}$.
Hence, when there is a Bose condensate, the chemical potential $\mu$ must
be equal to zero if we choose $E_{c}=0$.

We take the total energy of a set of charged bosons in localised
 states to be the sum of the energies of the bosons in the
individual potential wells.  The partition function $Z$ for such a system
is then
the product of the partition functions $z$ for each of the wells,
 and the system free energy $F=k_B T \ln Z$ is simply the
sum of the individual free energies $k_B T \ln z$.  Hence it makes sense to
study the
partition function for one
localised state of energy $E$  on its own. The free energy of
all localised bosons is then given by
\begin{equation}
F=k_B T\int_{-\infty}^{0}dE N_{L}(E)ln z(E),
\end{equation}
where $N_{L}(E)$ is the one-particle density of localised states
below the mobility edge.

There is an important reference case 
when the temperature dependence of the specific heat can be readily
established; this occurs when $\nu=1$ and $\mu=0$. In that case the
repulsive Coulomb
energy scales as $E$, Eq.(4). By introducing $E/k_B T$ as a new
variable in the integral, Eq.(5), and assuming that the density of states
$N_{L}$ is energy independent in the region of order $T/\kappa$ below the
mobility edge one obtains
\begin{equation}
C=-T{\partial^{2} F\over{\partial T^{2}}}\propto T.
\end{equation}
In the general case $\nu \neq 1$ and (or) $N_{L}$ not energy independent
 the specific heat has a non-linear temperature dependence.
In this paper we study the case of low doping when
$N_{L}(\epsilon)=n_{L}\delta(E-\epsilon)$ where $n_{L}$ is the number of
impurity wells with only one localised level in each of them.

\section{Charged Bosons in a Single Localised State}

We focus on the properties of a single localised state $\epsilon$.  The
probability for the state to contain $p$ bosons is proportional to
$e^{-\beta \left(w-p\mu\right)}$
where $\mu$ is the chemical potential and $\beta \equiv 1/k_{B}T$.  (We
shall retain the possibility
of a non-zero chemical potential until it starts to complicate the
equations.) We can re-express $w-p\mu$ as
\begin{equation}
 w-p\mu =
         \kappa\left(-\epsilon\right)^\nu\left(p-p_0\right)^2
                -\kappa\left(-\epsilon\right)^{\nu}p_{0}^{2}
\end{equation}
where
   \begin{equation}
      p_0 =  \frac{1}{2} + \frac{\mu-\epsilon}{2\kappa(-\epsilon)^\nu}
   \end{equation}
Fig 1 shows a graph of $(w-p\mu)/[\kappa(-\epsilon)^\nu]$ as a function of
$p$.  The partition
function $z(\epsilon)$ for such a single localised state is
\begin{eqnarray}
    z(\epsilon) & = & \sum_{p=0}^{\infty} e^{-\beta (w-p\mu)}
\label{e:zeta}\\
            & = & e^{p_{0}^{2}\beta \kappa\left(- \epsilon\right)^{\nu}}
       \sum_{p=0}^{\infty}
        e^{-\beta \kappa\left(-\epsilon\right)^{\nu}\left(p-p_0\right)^2}.
                \label{e:zls}
\end{eqnarray}
The partition function is thus completely determined by the dimensionless
parameters $p_0$ and $k_{B}T/[\kappa(-\epsilon)^{\nu}]$.  The mean occupancy
$\langle p \rangle$ is
\begin{equation}
        \langle p \rangle=k_{B}T\frac{\partial \ln
z(\epsilon)}{\partial\mu}  \label{e:<p>}
        \end{equation}
and, when $\mu=0$, the specific heat capacity $c$ is
\begin{equation}
        c = \beta^{2}\frac{\partial^2 \ln  z(\epsilon)}{\partial \beta^{2}}.
                        \label{e:c_v}
        \end{equation}
Truncating the series at 100 terms the calculated values of these
quantities are shown in Figs~2 and~3.  We now attempt to understand these
results in more detail, looking separately at each temperature range.

\begin{enumerate}
  \item $k_{B}T\ll\kappa\left(-\epsilon\right)^{\nu}$

At low temperatures the partition function is dominated by the term with
$p$ closest to $p_0$, i.e. the value of $p$ giving the lowest value of
$w-p\mu$, and so the mean occupancy $\langle p \rangle$ is an integer and
goes up in steps as $p_0$ increases, as seen in Fig~2a.  The changeover in
dominance from one term to another occurs when $p_0$ is a half-integer, at
which point the two lowest energy states are degenerate.

So long as one term dominates the partition function, the specific heat
capacity $c$ will be close to zero.  However when $p_0$ is close to a
half-integer we have a two level system and a corresponding Schottky
anomaly in the specific heat capacity.  This is seen in Fig~3a: $c$ is
zero when $p_0$ is equal to a half-integer and rises to a maximum on
either side, when the level separation is $\sim k_{B}T$.  Hence, at fixed
$k_B T/[\kappa(-\epsilon^\nu)]$,  the low temperature specific heat capacity
(i.e. $k_{B}T\ll p_0^2
\kappa(-\epsilon)^\nu$) is periodic in $p_0$. We also note from Fig.~3
that the peak in the specific heat capacity rises to a maximum when $p_0$
is an integer. This is because the lowest energy level is the only
non-degenerate level, all others being in degenerate pairs;  hence, at low
temperature, we effectively have a two level system in which the upper
level is a degenerate pair, thus resulting in a larger peak in $c$ than
occurs in an ordinary two level system.

  \item $k_{B}T>\kappa(-\epsilon)^{\nu}$

  We can approximate the sum by an integral
\begin{equation}
 z(\epsilon) \approx
e^{p_{0}^{2}\beta\kappa\left(-\epsilon\right)^{\nu}}
\int_{0}^{\infty}dp e^{-\beta \kappa(-\epsilon)^{\nu}(p-p_0)^2}
\end{equation}

    \begin{enumerate}
      \item $\kappa(-\epsilon)^{\nu}<k_{B}T<p_{0}^{2}\kappa(-\epsilon)^{\nu}$

In this case we can approximate the lower limit of the integral
as~$-\infty$, i.e. the partition function can be approximated by an
untruncated gaussian, and is therefore approximately symmetrical about
$p_0$. Hence, in this temperature range we have
  \begin{equation}
      \langle p \rangle \approx p_0
  \end{equation}
as is clearly seen in Figs~2a and~2b.

After integration, the partition function becomes
\begin{equation}
 z(\epsilon)\approx e^{\beta p_{0}^{2}(\kappa-\epsilon)^{\nu}}
        \sqrt{\frac{\pi k_{B}T}{\kappa(-\epsilon)^{\nu}}}
\end{equation}
from which we obtain that
\begin{equation}
  c\approx\frac{1}{2} k_B
\end{equation}
This result is another example of the equipartition theorem of classical
statistical mechanics and arises simply because $w$ is proportional to the
square of a co-ordinate, namely $p-p_0$.  In Fig.~3b the effect of this
result is seen in the region around $k_{B}T\approx\kappa(-\epsilon)^\nu$.

\vspace{5 mm}

      \item $k_{B}T>\kappa(-\epsilon)^{\nu}$ and $k_{B}T \gg
p_{0}^{2}\kappa(-\epsilon)^{\nu}$

In this case we can make the approximation that
\begin{equation}
  \int_{0}^{p_0}dp e^{-\beta \kappa(-\epsilon)^{\nu}(p-p_0)^{2}}
   \approx p_0
\end{equation}
and so  the partition function becomes
\begin{equation}
   z \approx p_0 + \frac{1}{2}
   \sqrt{\frac{\pi k_{B}T}{\kappa\left(-\epsilon\right)^{\nu}}}
\end{equation}
As a consequence, the mean occupancy $\langle p \rangle$ becomes
proportional to the square root of temperature:
   \begin{equation}
        \langle p \rangle \approx \sqrt{\frac{k_{B}T}{\pi\kappa(-\epsilon)^\nu}}
        \end{equation}
and this gives rise to the increase in $\langle p \rangle$ at large
temperatures seen in Fig.~2. Once again, the specific heat capacity
$c\approx\frac{1}{2}k_B$,
as seen in
Fig.~3.

 \end{enumerate}
    \end{enumerate}

\section{Bose-glass - superfluid transition}

Here we assume that all the localised states have the same value of
$\epsilon$ and derive the conditions for the Bose condensate to occur.
The most startling conclusion is that, in particular circumstances, it is
possible to take a system in which there is no Bose condensate and create
one by raising the temperature.

If the de-localised bosons are treated as being free particles of spin-0
and mass $m$, then at temperature $T$ and when $\mu=0$, the number of
de-localised bosons per unit cell of volume $\Omega$ is
\[ \left(\frac{T}{T_{c0}}\right)^{\frac{3}{2}}  \]
where
\begin{equation}
  k_{B}T_{c0}=3.3125\frac{\hbar^2}{m}
\left(\frac{1}{\Omega}\right)^{\frac{2}{3}}
\end{equation}
If there are $n_L$ localised states in each unit cell and the mean
occupancy of each state is $\langle p \rangle$ then the number of
localised bosons per unit cell is
\[  n_L \langle p \rangle \]
If there are $n$ bosons, in total, per unit cell, then the number of
bosons in the Bose condensate is
\[   n- n_L \langle p \rangle -
\left(\frac{T}{T_{c0}}\right)^{\frac{3}{2}} \]
and the temperature of the phase transition $T_{c}$ is found by solving
\begin{equation}
  n- n_L \langle p \rangle - \left(\frac{T_{c}}{T_{c0}}\right)^{\frac{3}{2}} =
0.
                \label{e:tc}
\end{equation}
  However, as we have seen, $\langle p \rangle$ is in general a
function of $T$ and so solving this equation is not trivial.
Nevertheless we can make some simple observations about the conditions
necessary for a Bose condensate to occur.  Fig.~4 summarises the
conclusions.

When $p_0$ is just above an integer value $p_-$, $\langle p \rangle$
increases monotonically with temperature (as in Fig.~2b); we then have two
possibilities:
\begin{enumerate}
  \item $n<n_{L}p_-$

  No Bose condensate occurs at any temperature.

  \item $n>n_{L}p_-$

  In this case equation Eq.(\ref{e:tc}) has one solution, being the temperature
above which the Bose condensate disappears.

\end{enumerate}

When $p_0$ is just below an integer value $p_+$, on the other hand,
$\langle p \rangle$ no longer increases monotonically with temperature
(again, see Fig.~2b); rather, as $T$ increases above zero $\langle p
\rangle$ falls from $p_+$ to $p_0$ before rising back up again.  Now we
can distinguish three cases:
\begin{enumerate}
  \item $n<n_{L}p_0$

  No Bose condensate occurs at any temperature.

  \item $n>n_L p_+$

  At $T=0$ the Bose condensate is present, but disappears as the
temperature is raised.

  \item $n_L p_0 < n < n_L p_+$
  This is more complicated.  At $T=0$ the Bose condensate is absent.  If
  \[k_{B} T_{c0} < \sim \frac{\kappa(-\epsilon)^\nu}{n_{L}^{\frac{2}{3}}}  \]
  then the Bose condensate will be absent at all temperatures.  If, on the
other hand,
  \[k_{B} T_{c0} > \sim \frac{\kappa(-\epsilon)^\nu}{n_{L}^{\frac{2}{3}}}  \]
  then a Bose condensate will appear at a temperature
$k_{B} T\sim\kappa(-\epsilon)^\nu$ and disappear as the temperature is raised
further.
\end{enumerate}

\section{Conclusion}

We have formulated the statistics of charged  bosons partly localised by
impurities. A simple form of the partition function is proposed by the use
of a
reasonable scaling of the Coulomb energy with the localisation length. A
 non-uniform dependence of the specific heat of the partly localised
charged
superfluid is found at low temperatures which strongly depends on the
 exponent $\nu$ of the localisation length. The Bose-glass -
superfluid transition is analysed (Fig.4) as a function of the $n/n_{L}$
ratio and
the Coulomb interaction as characterised by the parameter $p_{0}$. A
new phase is found, which is a Bose-glass at $T=0$ but a superfluid at
finite temperatures.

We believe that our findings are relevant for doped high-$T_{c}$ cuprates
having many properties reminiscent of the  charged
Bose-liquid\cite{alemot}. Because the level of doping of these
Mott-Hubbard insulators is high, one can
expect $N_{L}(E)$ to vary smoothly with energy $E$ (rather than having
spikes corresponding to discrete levels). In that case our approach leads to
a non-linear temperature dependence of the specific heat at low temperatures
if $\nu
\neq 1$.
From a preliminary analysis, the experimental observation of $C\propto
T^{1-\delta}$ in superconducting
$La_{2-x}Sr_{x}CuO_{4}$ \cite{Hussey} appears to be consistent with the
existence of bosons
partly localised by disorder.

\vspace{7 mm}
  Enlightening   discussions with N. Hussey,
 V. Kabanov,  F. Kusmartsev,  J. Samson, and K.R.A. Ziebeck are
highly appreciated.

\pagebreak


%
%
\pagebreak
\begin{figure}
\caption{Graph of $(w-p\mu)/[\kappa(-\epsilon)^\nu]$ against $p$.}
\label{fig1}
\end{figure}

\begin{figure}
\caption{The mean occupancy $\langle p\rangle$ as a function of $p_0$ and
$\log_{10}\left\{k_{B}T/[\kappa(-\epsilon)^\nu]\right\}$.  (a) The full
3-D plot. (b) $\langle p \rangle$ versus
$\log_{10}\left\{k_{B}T/[\kappa(-\epsilon)^\nu]\right\}$ for selected values of
$p_0$.}
\label{fig2}
\end{figure}

\begin{figure}
\caption{The specific heat capacity $c$ as a function of $p_0$ and
$\log_{10}\left\{k_{B}T/[\kappa(-\epsilon)^\nu]\right\}$.  (a) The full
3-D plot.  (b) $c$ versus
$\log_{10}\left\{k_{B}T/[\kappa(-\epsilon)^\nu]\right\}$ for selected values of
$p_0$: 3.0, 3.2, 3.3, 3.4, 3.48 and 3.5.  At low temperatures, $c$ is
approximately invariant under $p_0\rightarrow p_0\pm 1$ and under
$p_0\rightarrow n-p_0$ where $n$ is any integer greater than $p_0$.}
\label{fig3}
\end{figure}

\begin{figure}
\caption{Dependence of the possibility for Bose condensate formation on $n/n_L$
and $p_0$.  In the shaded regions, no Bose condensate can exist at $T=0$
but as the temperature is raised a condensate may form depending on the
value of $n_L^{\frac{2}{3}}k_{B}T_{c0}/[\kappa(-\epsilon)^\nu]$.}
\label{fig4}
\end{figure}

\end{document}